\documentclass[runningheads]{llncs}
\usepackage{microtype}
\usepackage{graphicx}
\usepackage{subfig}
\usepackage{booktabs} 
\usepackage{listings}
\usepackage{algorithm}
\usepackage{algorithmic}
\usepackage{float}
\usepackage{xcolor}

\definecolor{eclipseStrings}{RGB}{42,0.0,255}
\definecolor{eclipseKeywords}{RGB}{127,0,85}
\colorlet{numb}{magenta!60!black}

\usepackage{hyperref}

\begin{document}
\title{A Distributed Trust Framework for Privacy-Preserving Machine Learning}
%
%

\author{Will Abramson\inst{1} \and
Adam James Hall\inst{1} \and
Pavlos Papadopoulos\inst{1} \and
Nikolaos Pitropakis\inst{1} \and
William J Buchanan\inst{1}}

\authorrunning{Abramson et al.}
%
\institute{Blockpass Identity Lab, Edinburgh Napier University, Edinburgh, United Kingdom\\
\email{\{will.abramson,adam.hall,pavlos.papadopoulos,n.pitropakis,b.buchanan\}@napier.ac.uk}}
\maketitle              
\begin{abstract}
When training a machine learning model, it is standard procedure for the researcher to have full knowledge of both the data and model. However, this engenders a lack of trust between data owners and data scientists. Data owners are justifiably reluctant to relinquish control of private information to third parties. Privacy-preserving techniques distribute computation in order to ensure that data remains in the control of the owner while learning takes place. However, architectures distributed amongst multiple agents introduce an entirely new set of security and trust complications. These include data poisoning and model theft. This paper outlines a distributed infrastructure which is used to facilitate peer-to-peer trust between distributed agents; collaboratively performing a privacy-preserving workflow. Our outlined prototype sets industry gatekeepers and governance bodies as credential issuers. Before participating in the distributed learning workflow, malicious actors must first negotiate valid credentials. We detail a proof of concept using Hyperledger Aries, Decentralised Identifiers (DIDs) and Verifiable Credentials (VCs) to establish a distributed trust architecture during a privacy-preserving machine learning experiment. Specifically, we utilise secure and authenticated DID communication channels in order to facilitate a federated learning workflow related to mental health care data.

\end{abstract}
\section{Introduction}
\vspace{-10pt}
Machine learning (ML) is a powerful tool for extrapolating knowledge from complex data-sets. However, it can also represent several security risks concerning the data involved and how that model will be deployed \cite{munoz2017towards}. An organisation providing ML capabilities needs data to train, test and validate their algorithm. However, data owners tend to be wary of sharing data with third-party processors. This is due to the fact that once data is supplied, it is almost impossible to ensure that it will be used solely for the purposes which were originally intended. This lack of trust between data owners and processors is currently an impediment to the advances which can be achieved through the utilisation of big data techniques. This is particularly evident with private medical data, where competent clinical decision support systems can augment clinician-to-patient time efficiencies \cite{hall2016predicting,ahmad2019barriers}. In order to overcome this obstacle, new distributed and privacy-preserving ML infrastructures have been developed where the data no longer needs to be shared or even known to the Researcher in order to be learned upon. \cite{Ryffel2018}.

In a distributed environment of agents, establishing trust between these agents is crucial. Privacy-preserving methodologies are only successful if all parties participate in earnest. If we introduce a malicious Researcher, they may send a Trojan model which, instead of training, could store a carbon copy of the private data. Conversely, if we introduce a malicious actor in place of a data owner, they may be able to steal a copy of the model or poison it with bad data. In cases of model poisoning, malicious data is used to train a model in order to introduce a bias which supports some malicious motive. Once poisoned, maliciously trained models can be challenging to detect. The bias introduced by the malicious data has already been diffused into the model parameters. Once this has occurred, it is a non-trivial task to de-parrallelise this information.

If one cannot ensure trust between agents participating in a Federated Learning (FL) workflow, it opens the workflow up to malicious agents who may subvert its integrity through the exploitation of resources such as the data or the ML model used. In this work, we show how recent advances in digital identity technology can be utilised to define a trust framework for specific application domains. This is applied to FL in a healthcare scenario. This reduces the risk of malicious agents subverting the FL ML workflow. Specifically, the paper leverages: Decentralized Identifiers (DIDs) \cite{dids}; Verifiable Credentials (VCs) \cite{verifiable_creds}; and DID Communication, \cite{didcomm}.  Together, these allow entities to establish a secure, asynchronous digital connection between themselves. Trust is established across these connections through the mutual authentication of digitally signed attestations from trusted entities. The authentication mechanisms in this paper can be applied to any data collection, data processing or regulatory workflow. 

\vspace{2pt}
This paper contributes the following:
\vspace{-5pt}
\begin{itemize}
    \item We improve upon the current state-of-the-art with respect to the integration of authentication techniques in privacy-preserving workflows.
    \item  We enable stakeholders in the learning process to define and enforce a trust model for their domain through the utilisation of DID mechanisms.
    \item We apply a novel use of DIDs and VCs in order to perform mutual authentication for FL.
    \item We provide a threat model which can be used to quantify the threats faced by our infrastructure, namely vanilla FL.
    \item We specify a peer-to-peer architecture which can be used as an alternative to centralised trust architectures such as certificate authorities, and apply it within a health care trust infrastructure.
\end{itemize}


Section~\ref{litreview} provides the background knowledge and describes the related literature. Furthermore, Section~\ref{methodology} outlines our implementation overview, followed by Section~\ref{threatmodel}, where the threat model of our infrastructure is set, alongside with its scope. In Section~\ref{evaluation}, we provide an evaluation of our system, and conclude with Section~\ref{conclusion} that draws the conclusions and outlines approaches for future work.

\section{Background \& Related Work}
\label{litreview}
\vspace{-10pt}
ML is revolutionising how we deal with data. This is catalysed by hallmark innovations such as AlphaGo \cite{holcomb2018overview}. Attention has turned to attractive domains, such as healthcare \cite{wiens2017machine}, self-driving cars and smart city planning \cite{hashem2016role}. Ernst and Young estimate that NHS data is worth £9.6 Billion a year \cite{ey_nhs}. While this burgeoning application of data science has scope to benefit society, there are also emerging trust issues. The data-sets required to train these models are often highly sensitive, either containing personal data - such as data protected under the GDPR in the EU \cite{voigt2017eu} - or include business-critical information.

Additionally developing and understanding ML models is often a highly specialised skill. This generally means that two or more separate parties must collaborate to train an ML model. One side might have the expertise to develop a useful model, and the other around the data which they want to train the model, to solve a business problem.

\vspace{-8pt}
\subsection{Trust and the Data Industry}
\vspace{-8pt}


Trust is a complicated concept that is both domain and context-specific.
Trust is directional and asymmetric, reflecting that between two parties, the trust is independent for each party \cite{field_guide_trust}. Generally, trust can be defined as the willingness for one party to give control over something to another party, based on the belief that they will act in the interest of the former party \cite{hoffman2002conceptualization}. In economic terms, it is often thought of as a calculation of risk, with the understanding that risk can never be fully eliminated, just mitigated through mutual trust between parties \cite{keymolen2016trust}. The issue of trust is ever-present in the healthcare industry. Healthcare institutions collect vast amounts of personal medical information from patients in the process of their duties. This information can in turn be used to train an ML model. This could benefit society by enhancing the ability of clinicians to diagnose certain diseases.

DeepMind brought the debate around providing access to highly sensitive, and public, data-sets to private companies into the public sphere when they collaborated with the Royal Free London NHS Trust in 2015. This work outlined 'Streams', an application for the early detection of kidney failure \cite{powles2017google}. However, the project raised concerns surrounding privacy and trust. DeepMind received patient records from the Trust under a legal contract dictating how this data could be used. Later this was criticised as being vague and found to be illegal by the Information Commissioner's Office \cite{streams_ico}. Furthermore, DeepMind did not apply for regulatory approval through the research authorisation process to the Health Research Authority - a necessary step if they were to do any ML on the data. The team working on Streams has now joined Google, raising further concerns about the linkage of personal health data with Google's other records \cite{streams_google}.

While there was significant push back against the DeepMind/Royal Free collaboration, this has not prevented other research collaborations. This includes the automated analysis of retinal images \cite{de2016automated} and the segmentation of neck and head tumour volumes \cite{chu2016applying}. In both these scenarios, the appropriate authorisation from the Health Research Authority was obtained, and the usage of the data transferred was clearly defined and tightly constrained. 

\vspace{-10pt}
\subsection{Decentralised Identifiers (DIDs)}
\vspace{-8pt}
DIDs are tools which can be used to manage trust in a distributed or privacy-preserving environment. DIDs represent a new type of digital identifier currently being standardised in a World Wide Web Consortium (W3C) working group \cite{dids}. A DID persistently identifies a single entity that can self-authenticate as being in control of the identifier. This is different from other identifiers which rely on a trusted third party to attest to their control of an identifier. DIDs are typically stored on a decentralised storage system such as a distributed ledger, so, unlike other identifiers such as an email address, DIDs are under the sole control of the identity owner 


Any specific DID scheme that implements the DID specification must be resolvable to its respective document using the DID method defined by the scheme. Many different implementations of the DID specification exist which utilise different storage solutions. These include Ethereum, Sovrin, Bitcoin and IPFS; each with their own DID method for resolving DIDs specific to their system \cite{did_methods}. 


The goal of the DID specification is thus to ensure \emph{interoperability} across these different DID schemes such that, it is possible to understand how to resolve and interpret a DID no matter where the specific implementation originates from. However, not all DIDs need to be stored on a ledger; in fact, there are situations where doing so could compromise the privacy of an individual and breach data protection laws, such as with GDPR. Peer DIDs are one such implementation of the DID specification that does not require a storage system, and in this implementation DIDs and DID documents are generated by entities who then share them when establishing peer-to-peer connections. Each peer stores and maintains a record of the other peer's DID and DID Document \cite{peer_did}.
\vspace{-10pt}

\subsection{DID Communication (DIDComm)}
\vspace{-8pt}
DIDComm \cite{didcomm} is an asynchronous encrypted communication protocol that has been developed as part of the Hyperledger Aries project \cite{aries}. The protocol uses information within the DID Document, particularly the parties' public key and their endpoint for receiving messages, to send information with verifiable authenticity and integrity. The DIDComm protocol has now moved into a  standards-track run by the Decentralized Identity Foundation \cite{didcomm_dif}.




As defined in Algorithm \ref{Algorithm 1}, Alice first encrypts and signs a plaintext message for Bob. She then sends the signature and encrypted message to Bob's endpoint. Once the transaction has been received Bob can verify the integrity of the message, decrypt it and read the plaintext. All the information required for this interaction is contained within Bob and Alice's DID Documents. Examples of public-key encryption protocols include ElGamal \cite{elgamal1985public}, RSA \cite{rivest1978method} and elliptic curve based \cite{wohlwend2016elliptic}. Using this protocol both Bob and Alice are able to communicate securely and privately, over independent channels and verify the authenticity and integrity of the messages they receive.

\begin{algorithm}[h]
\caption{DID Communication Between Alice and Bob}

\label{Algorithm 1}
\begin{algorithmic}[1]
\STATE Alice has a private key $sk_a$ and a DID Document for Bob containing an endpoint ($endpoint_{bob}$) and a public key ($pk_b$).
\STATE Bob has a private key ($sk_b$) and a DID Document for Alice containing her public key ($pk_a$).
\STATE Alice encrypts plaintext message ($m$) using $pk_b$ and creates an encrypted message ($e_b$).
\STATE Alice signs $e_b$ using her private key ($sk_a$) and creates a signature ($\sigma$).
\STATE Alice sends $(e_b, \sigma)$ to $endpoint_{bob}$. 
\STATE Bob receives the message from Alice at $endpoint_{bob}$.
\STATE Bob verifies $\sigma$ using Alice's public key $pk_a$
\IF{Verify$(\sigma, e_b, pk_a) = 1$}
\STATE Bob decrypts $e_b$ using $sk_b$.
\STATE Bob reads the plaintext message ($m$) sent by Alice
\ENDIF
\end{algorithmic}
\end{algorithm}

\vspace{-8pt}

\subsection{Verifiable Credentials (VCs)}
\begin{figure}
    \centering
    \includegraphics[width=0.8\linewidth]{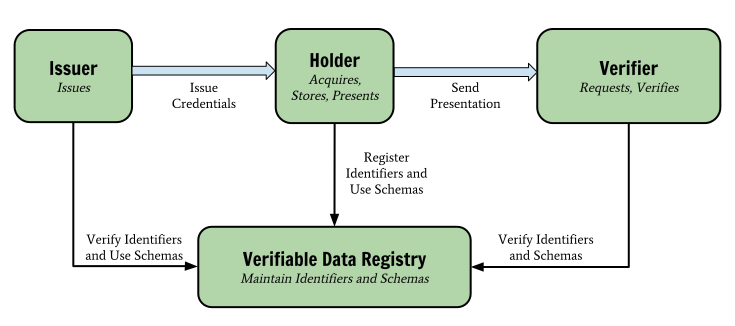}
    \caption{Verifiable Credential Roles \cite{verifiable_creds}}
    \label{fig:credential_roles}
\vspace{-18pt}
\end{figure}

\vspace{-6pt}
The Verifiable Credential Data Model specification became a W3C recommended standard in November 2019 \cite{verifiable_creds}. It defines a data model for a verifiable set of tamper-proof claims that is used by three roles; \emph{Issuer}, \emph{Holder} and \emph{Verifier} as it can be seen in Figure~\ref{fig:credential_roles}. A verifiable data registry, typically, a distributed ledger, is used to store the credential schemes, the DIDs, and DID documents of Issuers. 

\vspace{-12pt}

When issuing a credential, an Issuer creates a signature on a set of attributes for a given schema using a private key associated with their public DID through a DID document. The specification defines three signature schemes which are valid to use when issuing a credentials; JSON Web Signatures \cite{json_web_sigs}, Linked Data Signatures \cite{ld_sigs} and Camenisch-Lysyanskaya (CL) Signatures \cite{cl_signature_2003}. This paper focuses on the Hyperledger stack, which uses CL Signatures. In these signatures, a blinded link secret (a large private number contributed by the entity receiving the credential) is included in the attributes of the credential. This enables the credential to be tied to a particular entity without the Issuer needing to know the secret value.

When verifying the proof of a credential from a Holder, the Verifier needs to check a number of aspects:
\vspace{-4pt}
\begin{enumerate}
    \item The DID of the Issuer can be resolved on the public ledger to a DID document. This document should contain a public key which can be used to verify the integrity of the credential.
    \item The entity presenting the credential knows the link secret that was blindly signed by the Issuer. The Holder creates a zero-knowledge proof attesting to this.
    \item That the issuing DID had the authority to issue this kind of credential. The signature alone only proves integrity, but if the Verifier accepts credentials from any Issuers, it would be easy to obtain fraudulent credentials — anyone with a public DID could issue one. In a production system at-scale, this might be done through a registry, supported by a governance framework — a legal document outlining the operating parameters of the ecosystem \cite{rfcToIP}.
    \item The Issuer has not revoked the presented credential. This is done by checking that the hash of the credential is not present within a revocation registry (a cryptographic accumulator \cite{fischlin_dynamic_2009}) stored on the public ledger.
    \item Finally, the Verifier needs to check that the attributes in the valid credential meet the criteria for authorisation in the system. An often used example is that the attribute in your valid passport credential is over a certain age.
\end{enumerate}{}
\vspace{-4pt}
All the communications between either the Issuer and the Holder, or the Holder and the Verifier are done peer-to-peer using DIDComm. It is important to note that the Issuer and the Verifier never need to communicate.
\vspace{-8pt}
\subsection{Federated Machine-Learning (FL)}
\vspace{-6pt}
In a centralised machine learning scenario, data is sent to the Researcher, instead in a FL setup the model is being sent to each data participant.
The FL method has many variations, such as \emph{Vanilla}, \emph{Trusted Model Aggregator}, and \emph{Secure Multi-party Aggregation} \cite{DBLP:journals/corr/BonawitzIKMMPRS16}. However, at a high level, the Researcher copies one atomic model and distributes it to multiple hosts who have the data. The hosts train their respective models and then send the trained models back to the Researcher. This technique facilitates training on a large corpus of federated data \cite{Bonawitz2019} \cite{dean2012large}. These hosts then train models and aggregate these model updates into the final model. In the case of Vanilla FL, this is the extent of the protocol. However, we can extend this with a secure aggregator, a middle man in between the Researcher and the hosts, which averages participant models before they reach the Researcher. To further improve security, this can be extended using Secure Multiparty Aggregation to average models whilst they have been encrypted into multiple shares \cite{Ryffel2018}. The Researcher thus never sees the data directly, and only aggregates model gradients at the end \cite{das2016distributed}. However, this requires high network bandwidth and is vulnerable to invalidated input attacks \cite{bagdasaryan2018backdoor}, where an attacker might want to create a bias toward a particular classification type for a particular set of input data. 

\vspace{-12pt}
\section{Implementation Overview}
\label{methodology}
\begin{figure}[h!]
\centering
\includegraphics[width=0.8\linewidth]{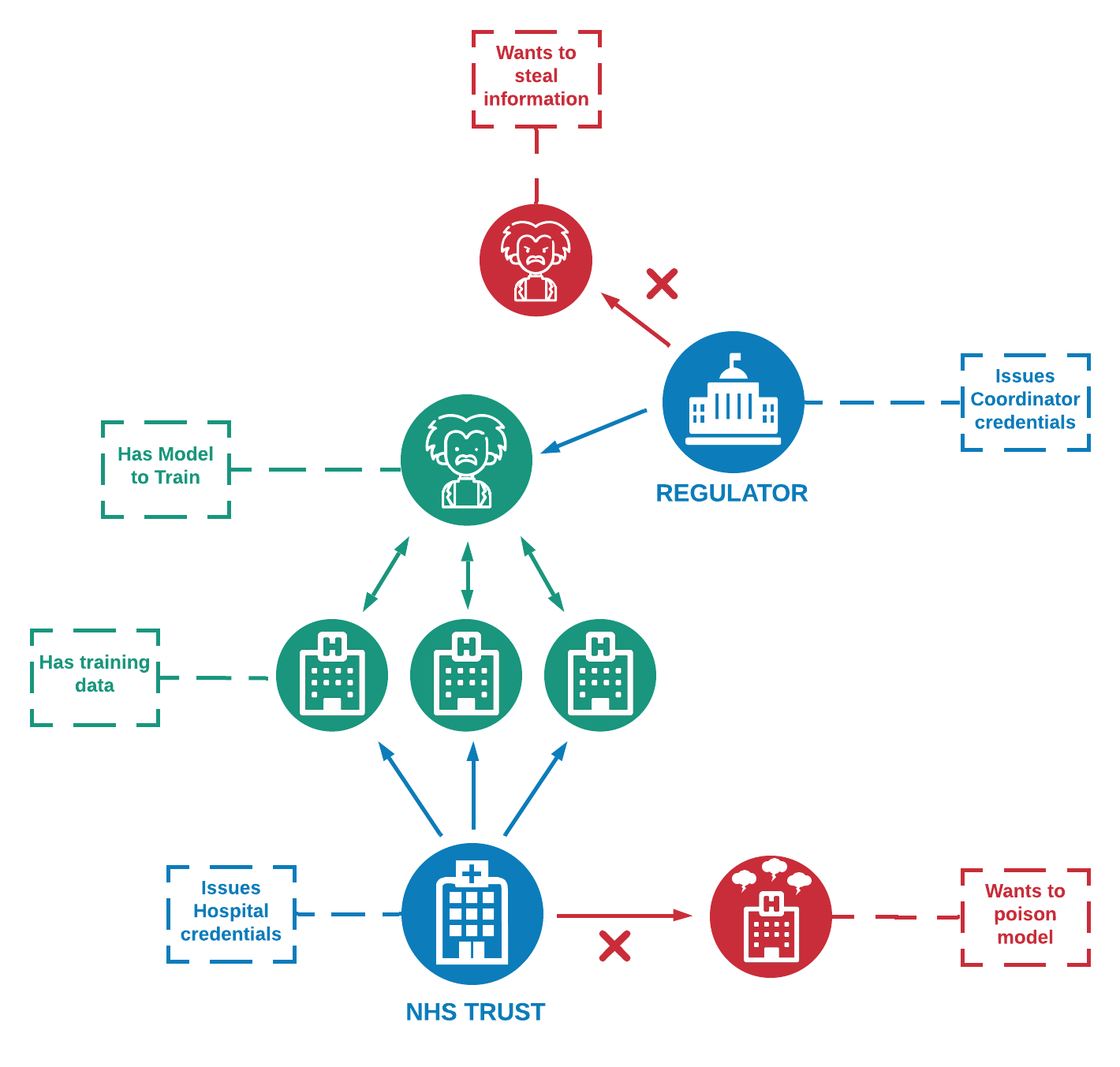}
\caption{ML Healthcare Trust Model}
\label{fighla_agents}
\vspace{-8pt}
\end{figure}

\vspace{-8pt}
Our work performs a basic FL example between Hyperledger Aries agents to validate whether distributed ML could take place over the DIDComm transport protocol. A number of Docker containers representing entities in a health care trust model were developed; creating a simple ecosystem of learning participants and trust providers (Figure \ref{fighla_agents}). For each Docker container running a Hyperledger Aries agent, we used the open-source Python Aries Cloud Agent developed by the Government of British Columbia \cite{aries_cloud_python}. Hospital containers are initialised with the private data that is used to train the model. 

\vspace{-12pt}

\vspace{-16pt}
\subsection{Establishing Trust}
\vspace{-8pt}
We define a domain-specific trust architecture using verifiable credentials issued by trusted parties for a healthcare use case. This includes the following agent types: a) NHS Trust (Hospital Credential Issuer); b) Regulator (Researcher Credential Issuer); c) Hospital (Data Provider); and d) Researcher (ML Coordinator).


This is used to facilitate the authorisation of training participants (verifiable Hospitals) and a Researcher-Coordinator. A data scientist who would like to train a model is given credentials by an industry watchdog, who in a real-world scenario could audit the model and research purpose. In the United Kingdom, for example, the Health Research Authority is well placed to fulfil this role. Meanwhile, Hospitals in possession of private health data are issued with credentials by an NHS authority enabling them to prove they are a real Hospital. The credential schema and the DIDs of credential Issuers are written to a public ledger — we used the development ledger provided by the government of British Columbia \cite{Britishcolumbia} for this work.

The system is established following the steps described in Algorithm~\ref{Algorithm:Trust}. Once both the \textit{Researcher-Coordinator} and the \textit{Hospital} agents have been authenticated, the communication of the model parameters for FL can take place across this secure trusted channel. 

\vspace{-12pt}
\begin{algorithm}[h!]
\caption{Establishing Trusted Connections}
\label{Algorithm:Trust}
\begin{algorithmic}[1]
\STATE \textit{Researcher-Coordinator} agent exchanges DIDs with the \textit{Regulator} agent to establish a DIDComm channel.
\STATE \textit{Regulator} offers an \textit{Audited Researcher-Coordinator} credential over this channel.
\STATE \textit{Researcher-Coordinator} accepts and stores the credential in their wallet.
\FOR{Each \textit{Hospital} agent}
\STATE Initiate DID Exchange with \textit{NHS Trust} agent to establish DIDComm channel.
\STATE \textit{NHS Trust} offers \textit{Verified Hospital} credentials over DIDComm.
\STATE \textit{Hospital} accepts and stores the credential.
\ENDFOR
\FOR{Each \textit{Hospital} agent}
\STATE \textit{Hospital} initiates DID Exchange with \textit{Researcher-Coordinator} to establish DIDComm channel.
\STATE \textit{Researcher-Coordinator} requests proof of \textit{Verified Hospital} credential issued and signed by the \textit{NHS Trust}.
\STATE \textit{Hospital} generates a valid proof from their \textit{Verified Hospital} credential and responds to the \textit{Researcher-Coordinator}.
\STATE \textit{Researcher-Coordinator} verifies the proof by first checking the DID against the known DID they have stored for the \textit{NHS Trust}, then \textit{resolving} the DID to locate the keys and verify the signature.
\IF{\textit{Hospital} can prove they have a valid \textit{Verified Hospital} credential}
\STATE \textit{Researcher-Coordinator} adds the connection identifier to their list of \textit{Trusted Connections}.
\ENDIF
\STATE \textit{Hospital} requests proof of \textit{Audited Researcher-Coordinator} credential from the \textit{Researcher-Coordinator}.
\STATE \textit{Researcher-Coordinator} uses \textit{Audited Researcher-Coordinator} credential to generate a valid proof and responds.
\STATE \textit{Hospital} verifies the proof, by checking the signature and DID of the Issuer.
\IF{\textit{Researcher-Coordinator} produces a valid proof of \textit{Audited Researcher-Coordinator}}
\STATE \textit{Hospital} saves connection identifier as a trusted connection.
\ENDIF
\ENDFOR
\end{algorithmic}
\end{algorithm}

\vspace{-28pt}
\subsection{Vanilla Federated Learning}
\vspace{-8pt}
This paper implements Federated Learning in its most basic form; where plain-text models are moved sequentially between agents. The Researcher-Coordinator entity begins with a model and a data-set to validate the initial performance. We train the model using sample public mental health data which is pre-processed into use-able training data. It is our intention to demonstrate that privacy-preserving ML workflows can be facilitated using this trust framework. Thus, the content of learning is not the focus of our work. 
We also provide performance results relating to the accuracy and resource-requirements of our system. We refer to our chosen workflow as Vanilla FL, and this is seen in Algorithm \ref{Algorithm:FL}. In order to implement Vanilla FL, the original data-set was split into four partitions, three training-sets and one validation-set. 



\vspace{-6pt}
\begin{algorithm}[h]
\caption{Vanilla Federated Learning}
\label{Algorithm:FL}
\begin{algorithmic}[1]
\STATE \textit{Researcher-Coordinator} has validation data and a \textit{model}, \textit{Hospitals} have \textit{training data}.
\WHILE{\textit{Hospitals} have unseen \textit{training data}}
\STATE \textit{Researcher-Coordinator} benchmarks \textit{model} performance against \textit{validation data} and sends \textit{model} to the next \textit{Hospital}.
\STATE This \textit{Hospital} trains the \textit{model} with their data and then sends the resulting \textit{model} back to the \textit{Researcher-Coordinator}.
\ENDWHILE
\STATE \textit{Researcher-Coordinator} benchmarks the final \textit{model} against \textit{validation} data.
\end{algorithmic}
\end{algorithm}
\vspace{-14pt}

This amalgamation of Aries and FL allowed us to mitigate some existing constraints centralised ML produced by a lack of trust between training participants. Specifically, these were: 1) Malicious data being provided by a false Hospital to spoil model accuracy on future cases; and 2) Malicious models being sent to Hospitals with the intention of later inverting them to leak information around training data values.
\vspace{-8pt}
\section{Threat Model}
\label{threatmodel}
\vspace{-10pt}
Since no data changes hands, FL is more private than traditional, centralised ML. However, some issues still exist with this approach. Vanilla FL is vulnerable to model stealing by ML data contributors who can store a copy of the Researcher's model after training it. In cases where the model represents private intellectual-property (IP), this setup is not ideal. On the other hand, with the knowledge of the model before and after training on each private data-set, the Researcher could infer the data used to train the model at each iteration \cite{shokri2017membership}. Model inversion attacks \cite{fredrikson2014privacy,fredrikson2015model} are also possible where, given carefully crafted input features and an infinite number of queries to the model, the Researcher could reverse engineer training values.

Vanilla FL is also potentially vulnerable to model poisoning and Trojan-backdoor attacks \cite{bagdasaryan2018backdoor,bhagoji2018analyzing,liu2017trojaning}. If data providers are malicious, it is possible to replace the original model with a malicious one and then send it to the Researcher. This malicious model could contain some backdoors, where the model will behave normally and react maliciously only to given trigger inputs. Unlike data poisoning attacks, model poisoning attacks remain hidden. They are more successful and easier to execute. Even if only one participant is malicious, the model's output will behave maliciously according to the injected poison. For the attacker to succeed, there is no need to access the training of the model; it is enough to retrain the original model with the new poisoned data. 

For the mitigation of the attacks mentioned above, our system implements a domain-specific trust framework using verifiable credentials. In this way, only verified participants get issued with credentials which they use to prove they are a trusted member of the learning process to the other entity across a secure channel. This framework does not prevent the types of attacks discussed from occurring, but, by modelling trust, it does reduce the risk that they will happen. Malicious entities could thus be checked on registration, or are removed from the trust infrastructure on bad behaviour.

Another threat to consider is the possibility of the agent servers or APIs getting hacked. Either the trusted Issuers could get compromised and issue credentials to entities that are malicious, or entities with valid credentials within the system could become corrupted. Both scenarios lead to a malicious participant having control of a valid verifiable credential for the system. This type of attack is a threat; however, it is outside the scope of this work. Standard cybersecurity procedures should be in-place within these systems that make successful security breaches unlikely. OWASP provides guidelines and secure practices to mitigate these \emph{traditional} cybersecurity threats \cite{owasp201810}. The defensive mechanisms are not limited to these and can be expanded using Intrusion Detection and Prevention Systems (IDPS) \cite{Modelvuln2019}.

\vspace{-10pt}
\section{Evaluation}
\label{evaluation}
\vspace{-10pt}
To evaluate the prototype, malicious agents were created to attempt to take part in the ML process by connecting to one of the trusted Hyperledger Aries agents. Any agent without the appropriate credentials, either a Verified Hospital or Audited Researcher-Coordinator credential, was unable to form authenticated channels with the trusted parties (Figure \ref{fighla_agents}). These connections and requests to initiate learning or contribute to training the model were rejected. Unauthorised entities were able to create self-signed credentials, but these credentials were rejected. This is because they had not been signed by an authorised and trusted authority whose DID was known by the entity requesting the proof.


The mechanism of using credentials to form mutually verifiable connections proves useful for ensuring only trusted entities can participate in a distributed ML environment. We note that this method is generic and can be adapted to the needs of any domain and context. Verifiable credentials enable ecosystems to specify meaning in a way that digital agents participating within that ecosystem can understand. We expect to see them used increasingly to define flexible, domain-specific trust. The scenario we created was used to highlight the potential of this combination. For these trust architectures to fit their intended ecosystems equitably, it is imperative to involve all key stakeholders in their design. 

Our work is focused on the application of a DID based infrastructure in a Federated Learning scenario. It is assumed that there is a pre-defined, governance-oriented trust model implemented such that key stakeholders have a DID written to an integrity assured ledger. The discovery of appropriate DIDs, and willing participants, either valid Researchers-Coordinators or Hospitals, related to a specific ecosystem is out-of-scope of our paper. This paper focuses on exploring how peer DID connections, once formed, facilitate participation in the established ecosystem. A further system can be developed for the secure distribution of the DIDs between the agents that are willing to participate.

\begin{figure}[h!]
    \centering
    \subfloat[CPU Usage (\%) during workflow]{{\includegraphics[width=5.65cm]{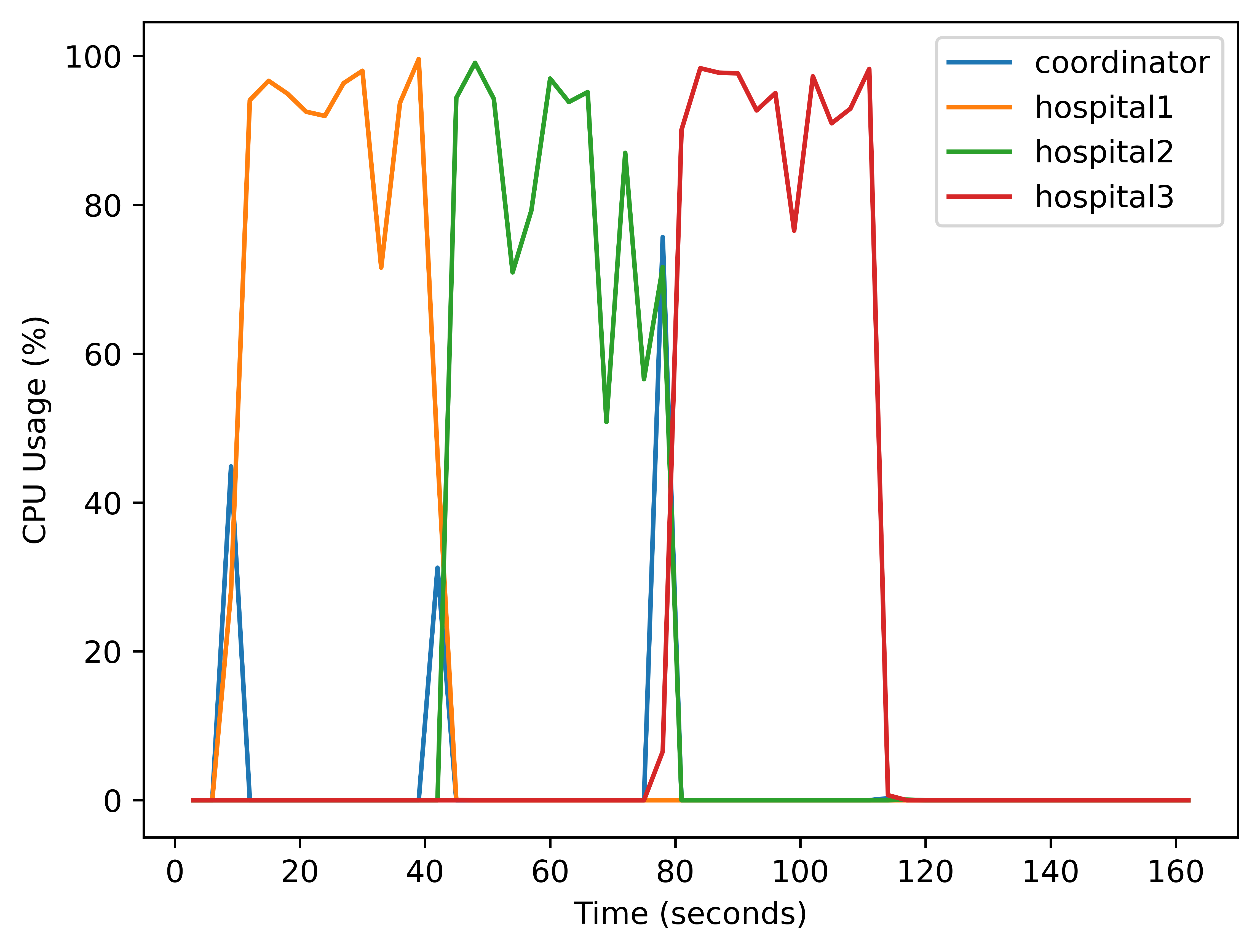} }}%
    \qquad
    \subfloat[Memory Usage (\%) during workflow]{{\includegraphics[width=5.65cm]{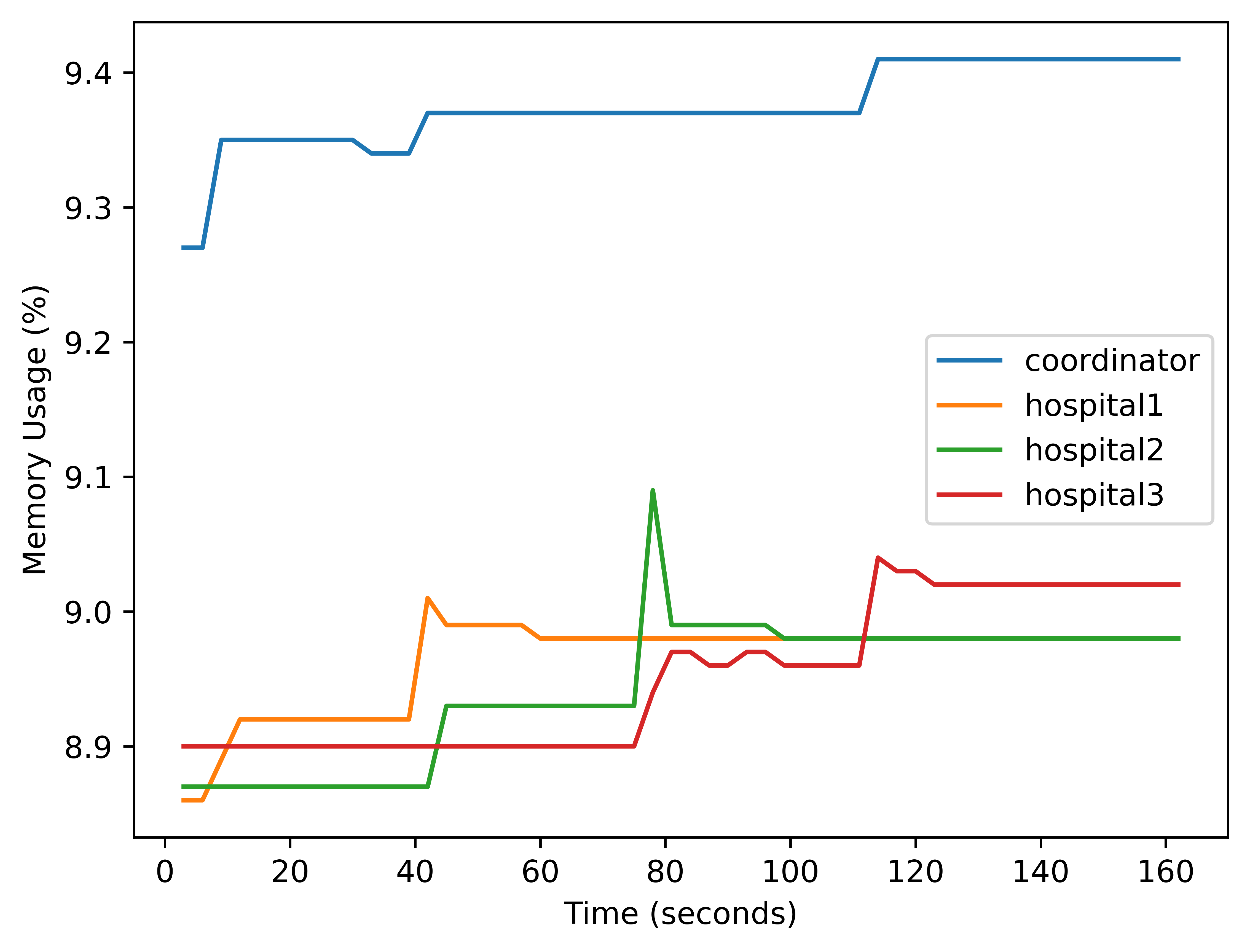} }}
    \qquad 
    \subfloat[Network Input (kB) during workflow]{{\includegraphics[width=5.65cm]{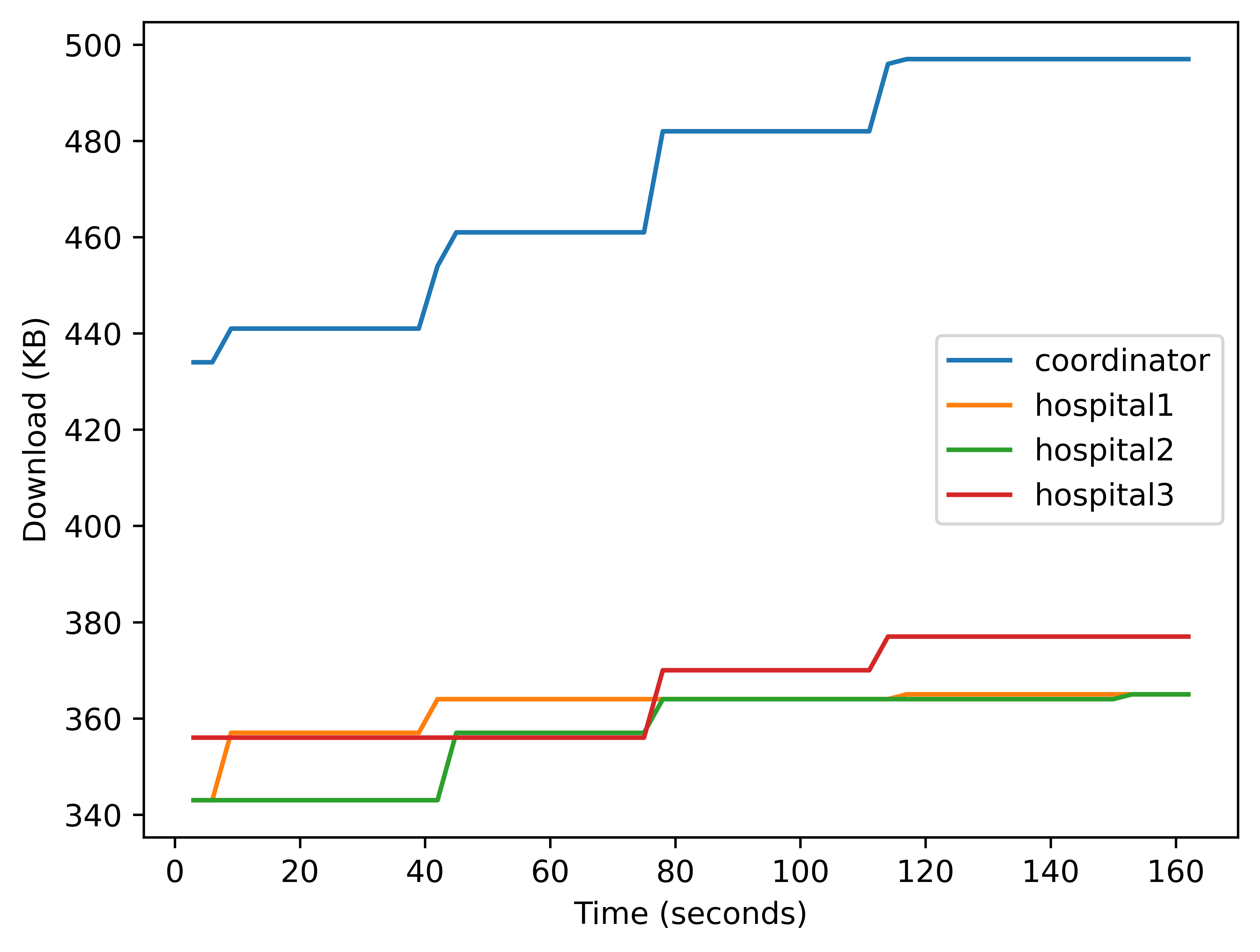} }}
    \qquad
    \subfloat[Network Output (kB) during workflow]{{\includegraphics[width=5.65cm]{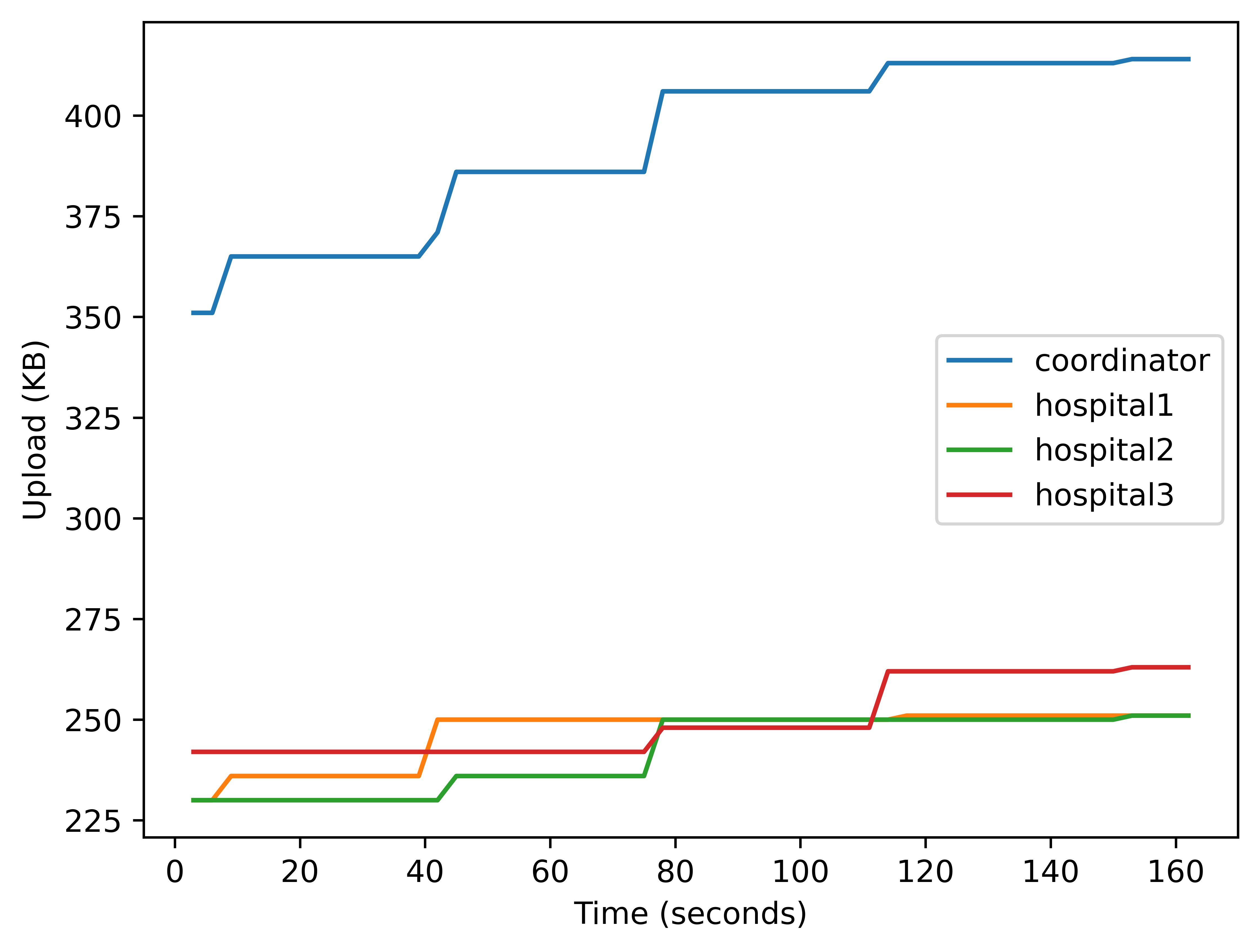} }}
    \caption{CPU, Memory Usage and Network Utilization of Docker container Agents during workflow}%
    \label{fig:networkoutput}%
    
\vspace{-0.2cm}
\end{figure}

Furthermore, performance metrics for each host were recorded during the running of our workflow. In Figure~\ref{fig:networkoutput} a), we see the CPU usage of each agent involved in the learning workflow. The CPU usage of the Researcher-Coordinator raises each time it sends the model to the Hospitals, and CPU usage of the Hospitals raises when they train the model with their private data. This result is consistent with what is expected given Algorithm \ref{Algorithm:FL} runs successfully. The memory and network bandwidths follow a similar pattern, as it can be seen in Figure~\ref{fig:networkoutput} b), Figure~\ref{fig:networkoutput} c) and Figure~\ref{fig:networkoutput} d). The main difference is that since the Researcher-Coordinator is averaging and validating each model against the training dataset every time, in turn, the memory and network bandwidth raises over time. From these results we can conclude that running federated learning in this way is compute heavy on the side of the Hospitals but more bandwidth and slightly more memory intensive on the side of the Researcher-Coordinator.
\vspace{-17pt}

The aim of this research is to demonstrate that a decentralised trust framework could be used to perform a privacy-preserving workflow. The authors train a dummy model on some basic example data. The intention here is merely to demonstrate that this is possible using our trust framework. 
We give the confusion matrix of the model tested on the Researcher-Coordinator's validation data after each federated training batch. This demonstrates that our model was successfully adjusted at each stage of training upon our federated, mental health dataset. The model develops a bias toward false-positives and tends to get less TNs as each batch continues. However, this may be due to the distribution of each data batch. Other than this, the learning over each batch tends to maximise true-positives. This can be observed in Table \ref{table:ConfMat}.
\vspace{-18pt}
\bgroup
\def\arraystretch{1.3} 
\begin{table}[h!]
\caption{Classifier's Accuracy Over Batches} 
\vspace{0.2cm}
\centering 
\begin{tabular}{c c c c c c c} 
\hline 
\textbf{Batch} & \textbf{0} & \textbf{1} & \textbf{2} & \textbf{3} \\ [0.5ex] 
\hline 
\textbf{True Positives} & 0 & 109 & 120 & 134\\
\textbf{False Positives} & 0 & 30 & 37 & 41 \\
\textbf{True Negatives} & 114 & 84 & 77 & 73\\
\textbf{False Negatives} & 144 & 35 & 24 & 10\\
\hline 
\end{tabular}
\label{table:ConfMat} 
\vspace{-30pt}
\end{table}
\egroup

\section{Conclusion \& Future Work}
\label{conclusion}
\vspace{-8pt}
This paper combines two fields of research, privacy-preserving ML and decentralised identity. Both have similar visions for a more trusted citizen-focused and privacy-respecting society. In this research, we show how developing a trust framework based on Decentralised Identifiers and Verifiable Credentials for ML scenarios that involve sensitive data can enable an increased trust between parties, while also reducing the liability of organisations with data.

It is possible to use these secure channels to obtain a digitally signed contract for training, or to manage pointer communications on remote data. While Vanilla FL is vulnerable to some attacks as described in Section~\ref{threatmodel}, the purpose of this work was to develop a proof of concept showing that domain-specific trust can be achieved over the same communication channels used for distributed ML. Future work includes integrating the Aries communication protocols, which enables the trust model demonstrated here, into an existing framework for facilitating distributed learning, such as PyGrid, the networking component of OpenMined \cite{Ryffel2018}. This will allow us and others to apply the trust framework to a far wider range of privacy-preserving workflows.

This will also allow us to enforce trust, mitigating model inversion attacks using differentially private training mechanisms \cite{dwork2011differential}. Multiple techniques can be implemented for training a differentially private model; such as PyVacy \cite{PyVacy} and LATENT \cite{chamikara2019local}. To minimise the threat of model stealing and training data inference, Secure Multiparty Computation (SMC) \cite{lindell2005secure} can be leveraged to split data and model parameters into shares. SMC allows both gradients and parameters to be computed and updated in a decentralised fashion while encrypted. In this case, custody of each data item is split into shares to be held by relevant participating entities. 

In our experiments, we utilised the Hyperledger Aries messaging functionality to convert the ML model into text and to be able to send it to the participating entities. In future work, we will going to focus on expanding the messaging functionality with a separate structure for ML communication. We also hope to evaluate the type of trust that can be placed in these messages, exploring the Message Trust Context object suggested in a Hyperledger Aries RFC \cite{aries_mtc}.

In this work, we address the issue of trust within the data industry. This radically decentralised trust infrastructure allows individuals to organise themselves and collaboratively learn from one another without any central authority figure. This breaks new ground by combining privacy-preserving ML techniques with a decentralised trust architecture.

\vspace{-8pt}

%
%
%
\bibliographystyle{splncs04}
\bibliography{references}
%




\end{document}